\documentclass[aip,reprint,english,a4paper] {revtex4-1}
\usepackage{upgreek}
\usepackage{amsfonts}
\usepackage{amstext}
\usepackage{mathtools}
\usepackage{amssymb}
\usepackage{color}
\usepackage{amsmath} 
\usepackage{graphicx} 

\newcommand{\bibdir}{Bib}
\newcommand{\dirfig}{Figures}

\newcommand{\SMTAna} {S1} 
\newcommand{\SMmultexp}{S2}
\newcommand{\SMmodel}{S3}
\newcommand{\SMacf}  {S4}
\newcommand{\SMAib}  {S5}

\begin{document}	

\title{Log-periodic oscillations as real-time signatures\\
of hierarchical dynamics in proteins}
\author{Emanuel Dorbath}
\affiliation{Biomolecular Dynamics, Institute of Physics, University
  of Freiburg, 79104 Freiburg, Germany}
\author{Adnan Gulzar}
\affiliation{Biomolecular Dynamics, Institute of Physics, University
  of Freiburg, 79104 Freiburg, Germany}
\author{Gerhard Stock}
\affiliation{Biomolecular Dynamics, Institute of Physics, University
  of Freiburg, 79104 Freiburg, Germany} 
\email{stock@physik.uni-freiburg.de}
\date{\today}

\begin{abstract}
  The time-dependent relaxation of a dynamical system may exhibit a
  power-law behavior that is superimposed by log-periodic
  oscillations.  Sornette [Phys.\ Rep.\ {\bf 297}, 239 (1998)] showed
  that this behavior can be explained by a discrete scale invariance
  of the system, which is associated with discrete and equidistant
  timescales on a logarithmic scale.  Examples include such diverse
  fields as financial crashes, random diffusion, and quantum
  topological materials. Recent time-resolved experiments and
  molecular dynamics simulations suggest that discrete scale
  invariance may also apply to hierarchical dynamics in proteins,
  where several fast local conformational changes are a prerequisite
  for a slow global transition to occur. Employing entropy-based
  timescale analysis and Markov state modeling to a simple
  one-dimensional hierarchical model and biomolecular simulation data,
  it is found that hierarchical systems quite generally give rise to
  logarithmically spaced discrete timescales. By introducing a
  one-dimensional reaction coordinate that collectively accounts for
  the hierarchically coupled degrees of freedom, the free energy
  landscape exhibits a characteristic staircase shape with two
  metastable end states, which causes the log-periodic time evolution
  of the system. The period of the log-oscillations reflects the
  effective roughness of the energy landscape, and can in simple cases
  be interpreted in terms of the barriers of the staircase landscape.
\end{abstract}
\maketitle

%
%
\section{Introduction}
\vspace*{-4mm}

\begin{figure}[ht!]
\centering
\vspace*{10mm}
\includegraphics[width=0.95\linewidth]{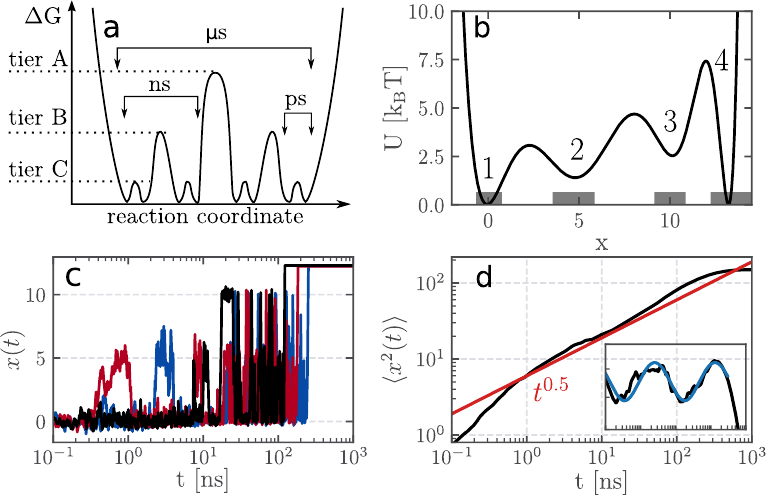}
\caption{
  Hierarchical dynamics in proteins. (a) Scheme of a hierarchical free
  energy landscape that represents dynamical processes on $\mu$s, ns
  and ps timescales by tiers A, B and C. (b) One-dimensional (1D)
  model potential $U(x)$, consisting of states ${\bf 1}-{\bf 4}$
  connected via energy barriers of similar height. Gray regions at the
  bottom define cores of the states used in Markov state modeling. (c)
  Time evolution $x(t)$ of three sample trajectories starting at
  $x=0$, which need several attempts to gradually climb over
  consecutive energy barriers. (d) Averaging over many trajectories,
  the resulting mean squared displacement ($\langle x^2(t)\rangle$, in
  black) shows a power-law behavior ($\propto t^{0.5}$, in red) that
  is superimposed by log-periodic oscillations. The enclosure shows a
  fit (blue) of the oscillatory residual $R(t)$ defined in Eq.\
  (\ref{eq:Rt}).}
\label{fig:model}
\end{figure}

Complex systems such as biomolecules exhibit motions on many
timescales, ranging from sub-picosecond vibrations to global
conformational rearrangements requiring seconds.\cite{Hu16,
  Lindorff-Larsen16} Rather than being uncoupled as assumed in normal
mode theory,\cite{Cui06} these molecular motions may interact in a
nonlinear and cooperative manner, such that fast fluctuations are a
prerequisite of rare transitions.\cite{Palmer84, Buchenberg15} The
basic concept of such a hierarchical coupling of multiscale motions is
often illustrated by a one-dimensional (1D) model of the free energy
landscape, which represents the dynamics on different timescales by
various tiers of the energy.\cite{Frauenfelder91, Dill97,
  Henzler-Wildman07a,Milanesi12} As an example, Fig.\ \ref{fig:model}a
displays a free energy landscape showing three tiers A, B and C, which
are associated with specific processes happening on $\mu$s, ns and ps
timescales, respectively. Because the system needs to cross the
barriers of tier C to reach the barriers of tier B and tier A, the
hierarchical model give a simple explanation of the mechanistic
coupling between fast and slow motions.

While the concept of a hierarchical energy landscape is appealing, the
microscopic nature of the tiers and the associated couplings between
them is not well understood. In principle, such mechanisms can be
inferred directly from all-atom molecular dynamics (MD) simulations,
which reveal the local structural changes that are required for a
global conformational rearrangement.\cite{Maisuradze13} For example,
by considering the left- to right-handed transitions of the helical
peptide Aib$_9$, Buchenberg et al.\cite{Buchenberg15} showed that
these global transitions (occurring on a $0.1 \mu$s timescale) first
require conformational transitions of individual residues (which take
about 1 ns), which in turn require the opening and closing of
structure-stabilizing hydrogen bonds (occurring within tens of ps).
Since the rates of these three processes were found to exhibit a
similar temperature behavior, they concluded that the heights of the
corresponding energy barriers must be similar, which appears to be in
contrast to the energy landscape shown in Fig.\ \ref{fig:model}a.

Represented by a one-dimensional (1D) model, these findings leads to a
staircase-like energy landscape $U(x)$ depicted in Fig.\
\ref{fig:model}b. The model consists of four states that are separated
by three energy barriers, whose similar height ($\sim 3\,k_{\rm B}T$)
corresponds roughly to the energy required to break a hydrogen
bond. In this way, $x$ could be considered as collective coordinate
constructed from the sum of three hydrogen bond
distances.\cite{Lickert21} Assuming that the system starts at time
$t = 0$ in state {\bf 1}, it evolves via intermediate states {\bf 2}
and {\bf 3}, and finally reaches the second low-energy state {\bf
  4}. Figure \ref{fig:model}c shows three sample trajectories $x(t)$
of the model, see Methods for details. Starting in state {\bf 1}, the
system will initially cross the first barrier to state {\bf 2}, but
then typically fall back to {\bf 1} due to the high back-rate of the
model. After some attempts, a rare fluctuation may drive the system
over the second barrier to reach state {\bf 3}, until after many
more attempts the system will eventually reach the final state {\bf
  4}. Using a logarithmic representation of time, this gradual
climbing over similar energy barriers manifest itself in apparent
log-oscillations of $x(t)$.

To explain these findings, we assume that the mean transition time over
the first barrier (with energy $E_b$) is given by
$\tau_{1\rightarrow 2} = \tau_0 e^{-\beta E_b}$, where $\beta = 1/k_{\rm
    B}T$ denotes the inverse temperature. Since all
barriers are similar, we roughly estimate that transitions over the
first two barriers take about
$\tau_{1\rightarrow 3} = \tau_0 e^{-2 \beta E_b}$, and
analogously $\tau_{1\rightarrow 4} = \tau_0 e^{-3\beta E_b}$.
This leads to the relation
\begin{equation} \label{eq:timescales}
\log \tau_{1 \rightarrow n+1} - \log \tau_{1 \rightarrow n} = \beta
E_b \log {\rm e},
\end{equation}
where $\log {\rm e} = \log x/\ln x $ denotes the decadic logarithm of
Euler's number. It states that the system exhibits discrete and
equidistant timescales on a logarithmic scale, which explain the
log-periodic oscillations of $x(t)$.

Log-periodic oscillations have been found in such diverse fields as
the diffusion on random lattices, \cite{Bernasconi82,Klafter91} in
dielectric relaxation \cite{Khamzin13} and the magnetoresistance of
ultraquantum topological materials, \cite{Wang18c} as well as in
large-scale phenomena such as earthquakes \cite{Sornette95} and
financial crashes. \cite{Johansen98,Geraskin13} Following
Sornette,\cite{Sornette98} they arise as a consequence of a discrete
scale invariance, meaning that the scale invariance exists only for
transformations $t \rightarrow \lambda t$ with discrete values of the
scaling parameter $\lambda $, that is,
$\lambda = \lambda_n = \kappa^n$. These relations results directly in
Eq.\ (\ref{eq:timescales}), when we set $\kappa = e^{\beta E_b}$ and
$\lambda_n = \tau_{1\rightarrow n}/\tau_0$.
Performing an ensemble average over many trajectories starting at
$x=0$, the theory predicts for the resulting mean position
$\langle x(t)\rangle$ and mean squared displacement
$\langle x^2(t)\rangle$ a power-law behavior superimposed by a weak
log-oscillatory pattern.\cite{Sornette98} Showing
$\langle x^2(t)\rangle$ together with the associated power law
$t^{0.5}$ and its residual oscillatory part, Fig.\ \ref{fig:model}d
reveals that this is indeed the case for the 1D model. While power
laws are the hallmark of scale-invariant and self-similar systems, the
emergence of log-periodic oscillations require the discreetness of the
underlying timescales.

Hence we have shown that the hierarchical structure of the 1D model
approximately obeys the condition of discrete scale invariance and
thus gives rise to log-periodic oscillations. While this condition is
indeed satisfied by various hierarchical models,
\cite{Bernasconi82,Sornette98,Metzler99, Lickert21} it is less clear
whether it is also obeyed by the free energy tiers of a real protein.
If so, log-periodic oscillations should be observable in MD
simulations as well as in time-resolved experiments.
This might be indeed the case. For example, Hamm and coworkers
designed photoswitchable proteins that initially trigger a local
photoinduced conformational change, whose propagation through the
protein can be monitored via transient infrared spectroscopy.
\cite{Buchli13,Bozovic20, Bozovic20a,Bozovic21,Bozovic22} Their
experiments on various PDZ domains exhibited strongly nonlinear
dynamics on timescales from picoseconds to tens of
microseconds. Accompanying MD studies of the nonequilibrium
conformational dynamics reproduced these findings and revealed a quite
complex structural reorganization of the protein
\cite{Nguyen06b,Buchenberg14, Buchenberg17,Ali22} In particular, the
time traces of both experiment and MD revealed overshootings, which
may indicate log-periodic oscillations.\cite{Stock18}

In this work we wish to explore the applicability and relevance of
discrete scale invariance for the modeling and understanding of
hierarchical dynamics in proteins. To this end, we first consider the
above 1D system as a proof-of-principle model. As we in general rely
on experimental or simulation results, we adopt a data-driven approach
that does not use information on the underlying theoretical
formulation of the model. Considering the time evolution of the system
as input data, we aim to construct a dynamical model of the underlying
dynamics. To focus first on ensemble-averaged data, we perform a
timescale analysis using a maximum entropy
method. \cite{Stock18,Lorenz-Fonfria06} Employing furthermore
single-molecule (i.e., single-trajectory) information,
\cite{Berezhkovskii20} we can recover the free-energy landscape of the
model and construct a Langevin equation\cite{Lange06b, Hegger09,
  Ayaz21} or a Markov state model. \cite{Bowman13a,Wang17a, Noe19} The
analyses are shown to result in a multiexponential response function
with discrete timescales, giving rise to log-periodic oscillations.

To test if the concepts are applicable to real data, we revisit the
above mentioned hierarchical dynamics of the achiral peptide helix
Aib$_9$.\cite{Buchenberg15} As the process evolves in a
high-dimensional coordinate space, we first need to define collective
variables that account for the hierarchically coupled degrees of
freedom. Employing nonequilibrium MD simulations, we again use a
timescale analysis and construct a Markov state model to identify the
timescales of the process. We study the conditions under which
log-periodic oscillations can be observed for Aib$_9$, and close with
a discussion what aspects of hierarchical dynamics may be learned from
these phenomena.

%
%
\section{Theory}
\vspace{-4mm}

As explained above, we wish to analyze a time series given from a
nonequilibrium experiment or MD simulation, using three theoretical
formulations: Maximum entropy timescale analysis,
\cite{Lorenz-Fonfria06} Markov state modeling,
\cite{Bowman13a,Wang17a,Noe19} and discrete scale
invariance.\cite{Sornette98} Moreover, we discuss if the same effects
can be also observed under equilibrium conditions.

\subsection{Timescale analysis}
\vspace{-4mm}

The time evolution of a relaxation process can
be described by a multiexponential response function
\begin{equation}\label{eq:Multiexp}
S(t) = s_0 - \sum_{k=1,K} s_{k} \,e^{-t/\widehat \tau_{k}} ,
\end{equation}
where $S(t)$ is the considered observable of the system (e.g.,
$S \!=\! \langle x \rangle$ for the 1D system), which was prepared at
$t\!=\!0$ in a nonequilibrium initial state (e.g., state {\bf 1} in
Fig.\ \ref{fig:model}b). The first term with
$\widehat \tau_0 = \infty$ gives the offset $s_0$, and the time
constants $\widehat \tau_k$ ($k \ge 1$) with amplitudes $s_k$ are
given in decreasing order.
To analyze the timescales inherent to a given time series $S(t)$,
we determine the timescale spectrum
$s(\widehat \tau_k) \equiv s_{k}$. To this end, we choose the time
constants to be equally distributed on a logarithmic scale (typically
10 terms per decade) and fit the corresponding amplitudes $s_k$ to
the data. When we assume that $S(t)$ is monotonic increasing, we can
choose $s_k \ge 0$. The local maxima of the spectrum 
\begin{equation}\label{eq:Multiexp2}
   \tau_n =  \{\tau,\; s(\tau) = {\rm max} \}
\end{equation}
are referred to as 'main timescales' of the process.

Equation (\ref{eq:Multiexp}) corresponds to an inverse Laplace
transformation, which is an ill-posed problem, because the included
exponential functions are not orthogonal to each other. To render the
fitting algorithm stable, we therefore introduce an entropy-based
regularization factor $S_{\rm ent}$ that enforces a smooth spectrum of
the amplitudes $s_k$. This is achieved by minimizing the weighted sum
$\chi^2 - \lambda_{\rm reg} S_{\rm ent} $ of this penalty function
together with the usual root mean square deviation $\chi^2$ of the fit
function to the data. \cite{Lorenz-Fonfria06} The regularization
parameter $\lambda_{\rm reg}$ controls whether the model is over- or
underfitted. $\lambda_{\rm reg}$ can be estimated via various
criteria; see the Supplementary Material for details (Fig.\ \SMTAna).

%
%
\subsection{Markov state modeling} \label{sec:MSM}
\vspace{-4mm}

If the free-energy landscape $\Delta G(x)$ of the system is known
(e.g., from single-trajectories), we may construct a Markov state
model (MSM), \cite{Bowman13a, Wang17a, Noe19} which describes
the dynamics in terms of memory-less jumps between $N$ metastable
conformational states of the system. Assuming a timescale separation
between fast intrastate fluctuations and rarely occurring interstate
transitions (i.e., the Markov approximation), the dynamics of the
system is completely determined by the transition matrix
$T(\tau_\text{lag})$ containing the probabilities $T_{ij}$ that the
system jumps from state $j$ to $i$ within lag time
$\tau_\text{lag}$. Denoting the state vector at time $t$ by
$\boldsymbol{p}(t) = (p_1,\ldots,p_N)^T$ with state probabilities
$p_i$, the time evolution of the MSM is given by
\begin{equation}\label{eq:CKtest}
\boldsymbol{p}(t\!=\!k\tau_{\rm lag})=T^k(\tau_{\rm lag})\boldsymbol{p}(0) .
\end{equation}
Upon diagonalizing the transition matrix $T$, we obtain its left/right
eigenvectors $\boldsymbol{\psi}^{\rm l}_n$/$\boldsymbol{\psi}^{\rm
r}_n$ and eigenvalues $\mu_n$. The latter yield the implied timescales
\begin{equation}\label{eq:imptime}
t_{n}= -\tau_\text{lag}/\ln \mu_{n}
\end{equation}
of the system, which correspond to experimentally measurable
quantities that account for the exponential decay $e^{-t/t_n}$
associated with eigenvector $\boldsymbol{\psi}_n$. Performing
an eigendecomposition of the transition matrix, the time
evolution can be written as a sum of decay factors multiplied with the
overlap of the eigenvectors with the initial state, \cite{Noe11}
\begin{equation}\label{eq:CKtest2}
\boldsymbol{p}(t\!=\!k\tau_{\rm lag})=\sum_{n=0,N-1}
\langle\boldsymbol{\psi}_n^{\rm r}|\boldsymbol{p}(0)\rangle\,  e^{-t/t_n} \,
\boldsymbol{\psi}_n^{\rm l} \, ,
\end{equation}
where $\boldsymbol{\psi}_0$ accounts for the equilibrium distribution
associated with $t_0=\infty$. Assuming that the observable of
interest, $S$, adopts in state {\bf i} the mean value
$\langle s \rangle_i$, we obtain again a multiexponential
representation of the time evolution of the observable, i.e.,
\begin{equation}\label{eq:CKtest3}
S(t) = \sum_{i=0,N-1} \langle s \rangle_i \, p_i(t) = \sum_{n=0,N-1}
\hat s_n \, e^{-t/t_n} 
\end{equation}
with $\hat s_n = \sum_i \langle s \rangle_i \langle {\bf i} |
\boldsymbol{\psi_n^{\rm l}}\rangle
\langle\boldsymbol{\psi}_n^{\rm r}|\boldsymbol{p}(0)\rangle$.\cite{Noe11}
While the MSM description of $S(t)$ has the same functional form as
the timescale analysis expression in Eq.\ (\ref{eq:Multiexp}), we note
that the sum now runs over the $N$ implied timescales $t_n$ of the
$N$-state system.

%
%
\subsection{Discrete scale invariance}
\vspace{-4mm}

Following Sornette,\cite{Sornette98} we summarize the basic ideas of
discrete scale invariance, which are relevant in the further discussion.
A time-dependent observable $S(t)$ is said to be {\em scale
  invariant} under the transformation $t \rightarrow \lambda t$, if
\begin{equation} \label{eq:SI} 
S(t) = \mu(\lambda) S(\lambda t) .
\end{equation}
For notational convenience, the time $t$ is given in dimensionless units.
The solution of this equation follows a power law with exponent
$\widetilde \alpha$,
\begin{equation} \label{eq:PL} 
S(t) = c t^{\widetilde \alpha} ,
\end{equation}
which can be verified by insertion. 
For example, the diffusion on a flat energy landscape ($U(x) = const.$)
is scale invariant, that is, it looks the same
on all time and length scales. 

While this is in general not true for diffusion on a
position-dependent potential $U(x)$, a weaker condition, {\em discrete
  scale invariance}, may apply if Eq.\ (\ref{eq:SI}) holds at least
for discrete values of the scaling parameter $\lambda$, that is,
\begin{equation} \label{eq:DSI} 
\lambda = \lambda_n = \kappa^n \quad (n \in \mathbb{N}) ,
\end{equation}
where $\kappa$ is the fundamental scaling ratio. This condition
reflects a symmetry of the problem, such as a lattice potential or the
above considered hierarchical landscape.
Inserting Eq.\ (\ref{eq:DSI}) in (\ref{eq:SI}), we obtain
\begin{equation} \label{eq:alpha1} 
\mu \lambda^{\widetilde \alpha} = 1 = e^{\mathrm{i} 2\pi m}  \quad (m \in \mathbb{Z}) ,
\end{equation}
where the second equation indicates that the exponent is in general
complex valued and can be written as
\begin{equation} \label{eq:alpha2} 
\widetilde \alpha = - \frac{\ln \mu}{\ln \lambda} + \mathrm{i} \frac{2\pi m}{\ln \lambda} 
\equiv \alpha + \mathrm{i} \omega .
\end{equation}
Using $t^{\mathrm{i} \omega} = e^{\mathrm{i}  \omega \ln t}$ and assuming that
$S = \mathrm{Re}\, S $, we obtain
\begin{equation} \label{eq:logOsc1} 
S(t) = c t^{\alpha} \cos (\omega \ln t) ,
\end{equation}
indicating that $S(t)$ exhibits log-periodic oscillations.
Assuming discrete scale invariance [Eq.\ (\ref{eq:DSI})], the
frequency $\omega$ depends only on the fundamental scaling ratio
$\kappa$, 
\begin{equation} \label{eq:DSI2} 
\omega = \frac{m 2\pi }{n \ln \kappa} = \frac{ 2\pi }{ \ln \kappa} \;,
\end{equation}
because $m$ is an arbitrary integer, which can be chosen as $m=n$.
This is in contrast to the case of general scale invariance (where
$\lambda$ can take an arbitrary
value), which results in a frequency $\omega \propto 1/\ln \lambda$
that is not constant. Hence, the existence of the log-oscillations
rests on the existence of discrete scaling parameters
$\lambda_n=\kappa^n$. 

Let us apply the above theory to our hierarchical model introduced in
Fig.\ \ref{fig:model}b. Showing consecutive barriers of similar height
$\beta E_b$, the system is expected to approximately exhibit
transition times,
$\tau_{1\rightarrow n} = \tau_0 e^{(n\!-\!1) \beta E_b}$, that are
discrete and equidistant on a logarithmic scale, see Eq.\
(\ref{eq:timescales}). Associating these timescales with the scaling
parameter via
\begin{equation} \label{eq:lambdan} 
\lambda_n = e^{n \beta E_b},
\end{equation}
we obtain for the fundamental scaling ratio
\begin{equation} \label{eq:kappa} 
\kappa = e^{\beta E_b},
\end{equation}
which is inverse proportional to the probability to cross single barrier.
Introducing $\tau_{\rm log} = \log {\rm e}\;2\pi / \omega$, we find
from Eqs.\ (\ref{eq:DSI2}) and (\ref{eq:kappa})
\begin{equation} \label{eq:logOsc2} 
 \tau_{\rm log} =  \beta E_b \log {\rm e}
\end{equation}
(with $\log {\rm e} \approx 0.43$), stating that the period of the
oscillations in decadic log time directly reflects the average barrier
height of the hierarchical energy landscape.

When we consider a time trace $S(t)$ obtained from an experiment or MD
simulation, we generally do not know the underlying energy landscape
and the corresponding transition times. Nonetheless, we may perform a
timescale analysis [Eq.\ (\ref{eq:Multiexp2})] to obtain the main
exponential timescales $\tau_1 \ge \tau_2 \ge \tau_3 \ldots$. In
direct analogy to Eq.\ (\ref{eq:timescales}), the existence of
log-oscillations then requires that these timescales are equally
spaced by their period $\tau_{\rm log}$,
\begin{equation} \label{eq:logOsc3} 
 \log  \tau_n - \log  \tau_{n+1} = \tau_{\rm log} \, .
\end{equation}
Note that this condition is certainly fulfilled in the common case
that only two main timescales exist, $\tau_1$ and $\tau_2$. When we
associate $\tau_2$ with the crossing of the first barrier $\beta E_1$
and $\tau_1$ with the timescale of the barrier $\beta E_{\rm tot}$ of
the overall transition, we obtain from our simple rate-theory
consideration
\begin{equation} \label{eq:logOsc2b} 
  \tau_{\rm log} =  \beta (E_{\rm tot} - E_1) \log {\rm e}
\end{equation}
stating that the log-period reflects the energy difference of the
overall and initial barriers.

Since the barrier heights of a 1D energy landscape do not necessarily
reflect the true reaction rates of a multidimensional
system,\cite{Altis08} in general we cannot directly associate
$\tau_{\rm log}$ with specific barriers of the system. Rather the
energy $\Delta E = \tau_{\rm log}/(\beta \log {\rm e})$ reflects an
overall roughness of the energy landscape, which gives rise to an
effective diffusion coefficient or intramolecular friction of the
process. \cite{Milanesi12,Zwanzig88,Schulz12,Echeverria14}

%
%
\subsection{Modeling log-periodic power laws} \label{sec:logPow}
\vspace{-4mm}

To be able to fit given MD data to a log-periodic power law $S(t)$, we
generalize Eq.\ (\ref{eq:logOsc1}) to the functional form
\begin{equation} \label{eq:logOsc4} 
S(t) = s_a + s_b t^{\alpha} + s_c t^{\alpha} \cos (
\tfrac{2\pi} {\tau_{\rm log}}  \log t + \varphi) ,
\end{equation}
which introduces the amplitudes $s_b$ and $s_c$ and the phase
$\varphi$ defining the oscillations, and the initial value $s_a$,
which is chosen such that $S(t) > 0$. 
To discuss the oscillations, we also consider the oscillatory
residual, defined as
\begin{equation} \label{eq:Rt} 
R(t) = (S(t) - s_a)t^{-\alpha} .
\end{equation}

It is instructive to study to what extent a simple multiexponential
model $S(t) = \sum_{n} s_{n} \,e^{-t/ \tau_{n}} $ with only two or
three main timescales $\tau_n$ and associated amplitudes $s_n$ can
give rise to the log-periodic power law in Eq.\ (\ref{eq:logOsc4}).
To this end, Fig.\ \SMmultexp{} shows power-law fits for models with
various choices of the timescales and amplitudes.  Using two
timescales that are more than one decade apart, i.e.,
$\log \tau_1 - \log \tau_2 \equiv \Delta_{12} \gtrsim 1$, we obtain
two well-defined log-oscillations with
$\tau_{\rm log} \approx \Delta_{12}$ for various choices of the
amplitudes. On the other hand, if the timescales are too close to each
other ($\Delta_{12} < 1$), we effectively see only a single
exponential term without log-oscillations. 
The scenario is similar for three timescales. If the timescale are
roughly logarithmically equidistant and well separated, i.e.,
$\Delta_{12} \approx \Delta_{23} \gtrsim 1$, we obtain three
well-defined log-oscillations. If two of the timescales are too close
to each other, we find only two exponential terms and two
log-oscillations.

%
%
\subsection{Nonequilibrium vs.\ equilibrium conditions} \label{sec:neq}
\vspace{-4mm}

In the discussion above, we have assumed nonequilibrium initial
conditions, i.e., we initially prepared the system in a nonstationary
state. This raises the question, if the same effects can be observed
under equilibrium conditions. Assuming linear response conditions, for
example, the nonequilibrium time evolution of the ensemble average
$\langle x(t)\rangle$ is known to be closely related to the
autocorrelation function
\begin{equation} \label{eq:acf} 
C(t) = \langle \delta x(t) \delta x(0) \rangle_{\rm eq} / \langle
\delta x^2 \rangle_{\rm eq}, 
\end{equation}
where $\delta x = x \!-\! \langle x \rangle$ and $\langle \ldots \rangle$
denotes a time average over an equilibrium trajectory.\cite{Chandler87}
In the nonequilibrium set-up considered in Fig.\ \ref{fig:model},
however, we typically stop the experiment or simulation, once the
system reaches the final state {\bf 4}. In this way, we prevent the
reverse reaction out of this state and the subsequent propagation
of the system, which would occur under equilibrium conditions. The
reverse reaction may introduce additional timescales that are not
encountered in the forward reaction. Moreover, the nonequilibrium
preparation results in well-defined initial conditions for the climb
over the energy landscape, which may lead to an oscillatory behavior of
the nonequilibrium ensemble average $\langle x(t)\rangle$. On the
other hand, the averaging over an equilibrium trajectory may average
over these oscillations.

%
%
\section{Discussion of the 1D model}
\subsection{Methods}
\vspace{-4mm}

Employing the 1D energy landscape $U(x)$ shown in Fig.\
\ref{fig:model}b, simulations of the Langevin equation
$m \ddot x(t)= - \frac{d U(x)}{dx} - \Gamma \dot x(t)+K\xi(t)$ were
performed as described in Ref.\ \onlinecite{Lickert21}. Specifically,
we used a mass of $m\!=\!1$\,u, constant friction
($\Gamma\!=\!130\,$kJ ps/mol) and noise amplitude
($K\!=\!25\,$kJ ps$^{1/2}$/mol), $\delta$-correlated Gaussian
noise $\xi$ of zero mean, and a temperature $T\!=\!300\,$K.
Collecting the data $x(t)$ every $\delta t = 0.4\,$ps, we transformed
to logarithmically spaced data for easier representation. To remove
fast fluctuations, a Gaussian filter with a standard deviation of 6
frames is used. Starting at $x\!=\!0$, in total $N\!=\!3000$
trajectories were run until they reach state {\bf 4}, i.e., for
$x\! \ge \!12.3$.
To calculate the mean position $\langle x(t)\rangle$, we performed an
ensemble average over the $N$ trajectories, where the initial
conditions of the individual trajectories were sampled from a
Boltzmann velocity distribution.
When we calculate the energy profile
$\Delta {\cal G} (x) = - k_{\rm B}T P(x)$ from the resulting
probability distribution $P(x)$, we consistently recover the potential
energy $U(x)$. \cite{note1}
Showing the convergence of the mean position $\langle x(t)\rangle$
with the sample size $N$, Figs.\ \SMmodel{a,b} reveal a
smooth increase of $\langle x(t)\rangle$ for $N \gtrsim 10^3$, while
for $N \lesssim 10^2$ we find additional high-frequency fluctuations
as signatures of transitions of individual trajectories. This
indicates that an incomplete sampling of the stochastic process can
be easily misinterpreted as log-periodic oscillations of
$\langle x(t)\rangle$.

%
%

As the 1D model in Fig.\ \ref{fig:model}b is well characterized in
terms of its states {\bf 1} to {\bf 4}, we may construct a MSM,
\cite{Bowman13a, Wang17a,Noe19} using our open-source Python
package msmhelper.\cite{Nagel23a} To this end, we first apply a
Gaussian smoothing with a standard deviation of 2.4\,ps to the
individual trajectories $x(t)$.\cite{Nagel23b} To avoid problems at
the boundaries, we define the states via their cores
\cite{Buchete08,Lemke16,Nagel19} (gray regions in Fig.\
\ref{fig:model}b), that is, we assign states {\bf 1} to {\bf 4} to the
$x$-regions [-0.6, 0.7], [3.6, 5.8], [9.2, 10.8], and
$x\! \ge \!12.3$. Intermediate frames between these regions are
assigned to the last populated state.  The resulting state
trajectories are then directly employed to calculate the transition
matrix $T(\tau_\text{lag})$.
Due to the coring procedure and since the underlying dynamics was
generated by a Markovian Langevin equation, we
expect the Markov approximation to hold well. As a consequence, the
implied timescales of the 4-state model are constant already at lag
time $\tau_\text{lag} = \delta t$, and the resulting Chapman-Kolmogorov
equation (\ref{eq:CKtest}) holds accurately, see Figs.\ \SMmodel{c,d}.

\begin{figure}[ht!]
\centering
\includegraphics[width=0.95\linewidth]{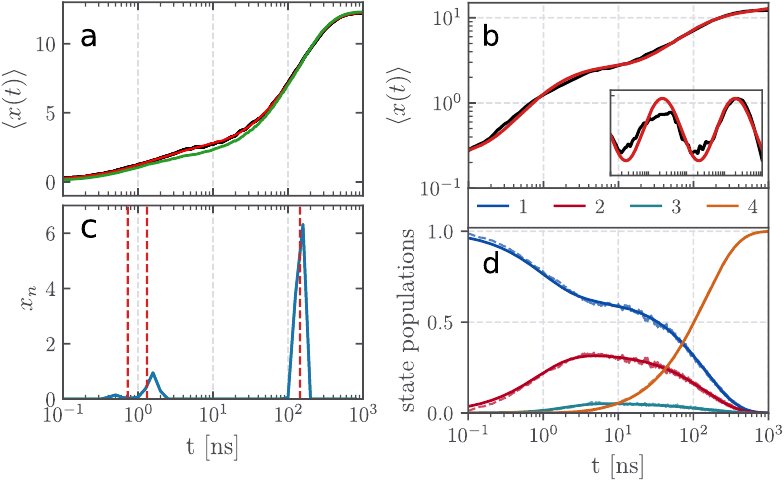}
\caption{\baselineskip4mm Analysis of the 1D model shown in Fig.\
  \ref{fig:model}b. (a) Mean position $\langle x(t)\rangle$, as
  obtained from (black) the Langevin simulations, (red) the fit of the
  timescale analysis [(Eq.\ (\ref{eq:Multiexp})], and (green) the MSM
  [(Eq.\ (\ref{eq:CKtest3})]. (b) Double-logarithmic representation of
  $\langle x(t)\rangle$ (black) together with the fit to Eq.\
  (\ref{eq:logOsc3}) (red), showing the oscillatory residual $R(t)$
  [Eq.\ (\ref{eq:Rt})] in the enclosure. (c) Timescale spectrum [Eq.\
  (\ref{eq:Multiexp}), in blue], compared to the implied timescales
  [Eq.\ (\ref{eq:imptime}), in red] of an MSM constructed from the
  Langevin data. (d) Time evolution of the state probabilities
  $p_n(t)$ of the MSM (thick lines) and the Langevin data (thin dashed
  lines).}
\label{fig:1danalysis}
\end{figure}

%
%
\subsection{Results}
\vspace{-4mm}

As detailed above, we run $N\!=\!3000$ Langevin simulations that start at
time $t\!=\!0$ in state {\bf 1}, and follow them until they reach the
final state {\bf 4}. Figures \ref{fig:1danalysis}a,b show the time
evolution of the resulting mean position $\langle x(t)\rangle$ in
log-time and double-log representation, which reflects the gradual
climb over the staircase-shaped energy landscape.
To identify the timescales of this climb, we model the time evolution
of $\langle x(t)\rangle$ by the multiexponential response function in
Eq.\ (\ref{eq:Multiexp}), see Methods. The resulting maximum-entropy
timescale spectrum ($\lambda_{\rm reg}\!=\!10$) shown in Fig.\
\ref{fig:1danalysis}c exhibits two main timescales at
$ \tau_1\! =\!160$\,ns and $ \tau_2 \!=\!1.6$\,ns, and a rather weak signature
at 0.5\,ns. From the time evolution of the individual trajectories
(Fig.\ \ref{fig:model}c), we conclude that the short timescale
reflects first attempts to cross the barrier to state {\bf 2}, while
the long timescale accounts for transitions to the final state {\bf
  4}.

%
%
%
From a data-driven view, the above timescale analysis requires only
ensemble-averaged data [e.g.,
$\langle x(t) \rangle \propto \sum_n x_n(t)$]. If single-trajectory
information [i.e., $x_n(t)$] is available,\cite{Berezhkovskii20} we
can calculate the probability distribution along $x$ and thus recover
the energy landscape $\Delta {\cal G}(x) = - k_{\rm B}T \ln P(x)$ of the
model. By identifying the four metastable conformational states {\bf
  1} to {\bf 4} of the system, we may then construct a MSM of the
dynamics (see Methods). As expected for a simple 1D system, we find
that the resulting implied timescales of the MSM, $t_1\!=\!145\,$ns,
$t_2\!=\!1.3\,$ns and $t_3\!=\!0.7\,$ns, agree well with the main
peaks of the timescale spectrum in Fig.\ \ref{fig:1danalysis}c.

What is more, the MSM provides an explanation of these timescale in
terms of the population flux between the states. Showing the time
evolution of the population probabilities $p_i(t)$ of the four states,
Fig.\ \ref{fig:1danalysis}d reveals that the population of initial
state {\bf 1} decays (within all three timescales $t_3$, $t_2$ and
$t_1$) such that the intermediate states {\bf 2} and {\bf 3} are
transiently populated (within $t_3$ and $t_2$), until the system
relaxes in final state {\bf 4} (within $t_1$). When we compare the
time evolution of the state populations obtained from the MSM to the
Langevin results, we find excellent agreement. Moreover, the MSM
calculation of the mean position $\langle x(t)\rangle$ via Eq.\
(\ref{eq:CKtest3}) matches perfectly the reference results (Fig.\
\ref{fig:1danalysis}a).

Interestingly, we find that the fastest timescale $t_3\!=\!0.7\,$ns
clearly shows up in the initial decay of $p_1(t)$ and the
corresponding initial rise of $p_2(t)$, although it is hardly visible
in the timescale analysis in Fig.\ \ref{fig:1danalysis}c. This
reflects the fact that the MSM exploits the structure of the energy
landscape, while the timescale analysis only uses the evolution of the
observable $\langle x(t) \rangle$, and thus depends considerably on
the definition of the collective coordinate $x$.

%

We are now in a position to assess if the dynamics of the 1D problem
fulfills the conditions of discrete scale invariance and therefore
gives rise to a log-periodic power law. 
Figure \ref{fig:1danalysis}c shows that both timescale
analysis and MSM essentially yield only two timescales,
$160$\,ns and $1.6$\,ns. (The third
MSM timescale carries hardly any amplitude and can be omitted.)
Alternatively, we may calculate the average transition times
$\tau_{1\rightarrow n}$ to reach state $n$ after starting in
state {\bf 1}, yielding $\tau_{1\rightarrow2}=2.6$\,ns,
$\tau_{1\rightarrow3}=23$\,ns and $\tau_{1\rightarrow4}=147$\,ns. As
$\tau_{1\rightarrow3}$ can be neglected (because state {\bf 3} is
hardly populated, see Fig.\ \ref{fig:1danalysis}d), we are again left
with two times, which resemble closely the timescales
determined by timescale analysis and the MSM. Because Eq.\
(\ref{eq:logOsc3}) is directly fulfilled if only two timescales
exist, the 1D problem is expected to show typical phenomena associated
with discrete scale invariance.

Using $ \tau_1 \!=\! 160$\,ns and $ \tau_2 \!=\! 1.6$\,ns, Eq.\
(\ref{eq:logOsc3}) predicts a log-periodic power law with an a period
of $\tau_{\rm log} \!=\! 2.0$. To test these predictions, we use the
functional form in Eq.\ (\ref{eq:logOsc4}) and fit the Langevin data to a
log-periodic power law. Figure \ref{fig:1danalysis}b shows that we
obtain a perfect fit, when we use $ \tau_{\rm log} \!=\! 1.92$ for the
period, $\alpha \!=\! 0.31$ for the exponent, and the coefficients
$s_a \!=\!-0.5$, $s_b \!=\! 1.7$, $s_c \!=\! -0.2$, and
$\varphi \!=\! 1.9$. The good agreement of theoretical [Eq.\
(\ref{eq:logOsc3})] and fitted values for $ \tau_{\rm log}$ is also
reflected in the fact that the maxima of the log-periodic oscillation
coincide well with the peaks of the timescale spectrum in Fig.\
\ref{fig:1danalysis}c.

Moreover, Eq.\ (\ref{eq:logOsc2b}) allows us to infer from the period
$\tau_{\rm log} \!=\! 2$ an effective roughness
$\beta \Delta E \!=\!4.7$ of the underlying energy landscape. In the
case of only two timescales, $\beta \Delta E$ can be interpreted as
the energy difference of the the overall and initial barriers. Indeed
the energy difference $\beta \Delta E = 4.5$ obtained from the
potential $U(x)$ in Fig.\ \ref{fig:model}b agrees well with the
theoretical prediction.
When we employ the simple rate approximation of Eq.\
(\ref{eq:lambdan}), we may also relate $\tau_{\rm log}$ to the average
barrier height $\beta E_b$ of the hierarchical energy landscape via
Eq.\ (\ref{eq:logOsc2}). In fact, $\beta E_b= 3.8$ obtained from the
potential $U(x)$ in Fig.\ \ref{fig:model}b matches
$\beta \Delta E \!=\!4.7$ at least qualitatively.

Apart from considering the mean $\langle x(t)\rangle$, it is
interesting to discuss the mean squared displacement
$\langle x^2(t)\rangle$, which accounts for the diffusional motion of
the system. As shown in Fig.\ \ref{fig:model}d, we obtain similar
log-oscillations with $ \tau_{\rm log} = 1.72$ and an increase of the
power-law exponent to $\alpha = 0.5$. That is, the hierarchical
dynamic of the 1D model manifest itself in subdiffusion, which
reflects the roughness of the underlying energy
landscape.\cite{Hu16,Milanesi12}

%
%

So far, we have assumed a nonequilibrium preparation of the system in
the initial state {\bf 1}. To test if the same effects can be observed
under equilibrium conditions, we run a $16\,\upmu$s-long equilibrium
simulation of the 1D model and calculate the autocorrelation function
$C(t)$ in Eq.\ (\ref{eq:acf}). Figure \SMacf{} shows that $C(t)$
decays on the timescales $\tau_1 = 26\,$ns and $\tau_2 =
1.3\,$ns. While $\tau_2$ is similar to the implied timescale $t_2$
found in the nonequilibrium case, $\tau_1$ is significantly shorter
than the nonequilibrium timescale $t_1 = 160\,$ns, which reflects the
possibility of back-reaction from state {\bf 4}. Moreover, we find
that the equilibrium autocorrelation function shows no evidence of
log-periodic oscillations. This indicates that the discrete scale
invariance of the observable $\langle x(t)\rangle$ requires
well-defined initial conditions (as given by a nonequilibrium
preparation) as well as a well-defined end state, which prevent the
averaging over the oscillations.

%
%
\section{Hierarchical dynamics of a peptide helix}

\begin{figure*}
\centering
\includegraphics[width=0.96\linewidth]{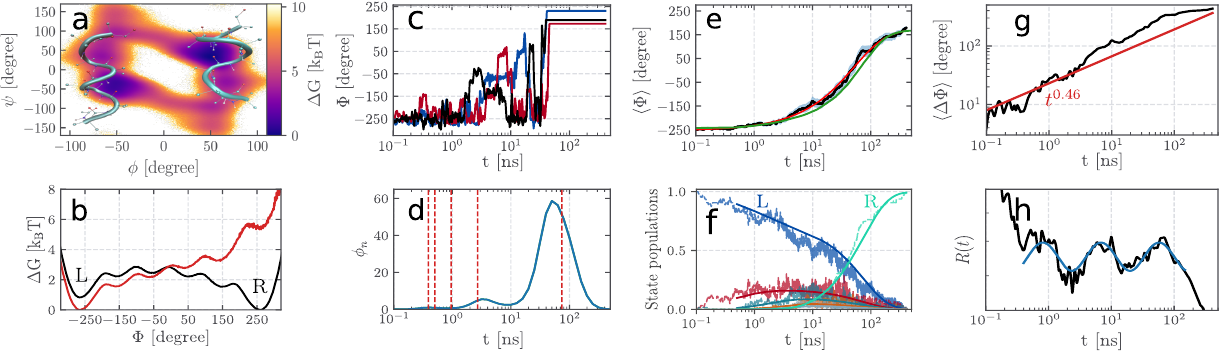}
\caption{\baselineskip4mm Hierarchical dynamics of the peptide helix
  Aib$_9$. (a) Structure of the left-handed and right-handed
  conformation of Aib$_9$, along with the Ramachandran plot
  $\Delta G(\phi,\psi)$, averaged over the five inner peptide
  residues. (b) Energy landscapes along the collective coordinate
  $\Phi$ [Eq.\ (\ref{eq:Phi})], obtained from equilibrium MD
  simulations\cite{Buchenberg15} (black) and from nonequilibrium
  trajectories that exhibit a single L$\rightarrow$R transition
  (red).\cite{note3} (c) Time evolution of three individual
  nonequilibrium trajectories $\Phi(t)$. (d) Timescale spectrum [Eq.\
  (\ref{eq:Multiexp}), in blue], compared to the implied timescales
  [Eq.\ (\ref{eq:imptime}), in red] of an MSM constructed from the MD
  data. (e) Mean position $\langle \Phi(t)\rangle$, as obtained
  from (black) the nonequilibrium MD trajectories, (red) the fit of
  the timescale analysis [(Eq.\ (\ref{eq:Multiexp})], and (green) the
  MSM [(Eq.\ (\ref{eq:CKtest3})]. (f) Time evolution of the state
  probabilities $p_n(t)$ of the MSM. (g) Double-logarithmic
  representation of
  $\langle \Delta \Phi(t)\rangle = \langle \Phi(t)\rangle + 255^\circ$
  together with the associated power law $t^{0.46}$ and (h) the
  residual oscillatory part $R(t)$ [Eq.\ (\ref{eq:Rt})].}
\label{fig:Aib}
\end{figure*}

It is interesting to study to what extent the theoretical concepts
established above for the 1D model can be transferred to the analysis of
all-atom MD data. As a well-established model system,
\cite{Buchenberg15,Perez18, Biswas18,Mehdi22} we consider the achiral
peptide Aib$_9$ that undergoes complete left- to right-handed chiral
transitions of the helix, see Fig.\ \ref{fig:Aib}a.

%
%
\subsection{Model and methods}
\vspace{-4mm}

Buchenberg et al.\cite{Buchenberg15} performed extensive MD
simulations of Aib$_9$
(H$_3$C-CO-(NH-C$_\alpha$(CH$_3$)$_2$-CO)$_9$-CH$_3$), using the
GROMACS program suite \cite{GROMACS05} with the GROMOS96 43a1 force
field\cite{GROMOS96} and explicit chloroform solvent.\cite{Tironi94}
Here we use eight of their MD trajectories at 320 K of each $2\ \mu$s
length, using a time step $\Delta t=$ 1 ps.
%
The resulting Ramachandran plot $\Delta G(\phi_i,\psi_i)$ along the
backbone dihedral angles $(\phi_i,\psi_i)$ of the five inner peptide
residues ($i=3,\ldots,7$) reveals a point symmetry with respect to
(0,0), which shows that Aib$_9$ indeed samples both left-handed
($\phi_i \ge 0$) and right-handed ($\phi_i \le 0$) conformations with
similar probability (Fig.\ \ref{fig:Aib}a). The corresponding two
local conformational states l at $\approx$ (-50$^\circ$, -45$^\circ$)
and r at $(50^\circ, 45^\circ)$ correspond to a right- and left-handed
helix, respectively.\cite{note3}
The ring-shaped free energy landscape $\Delta G$ reveals that left- to
right-handed transitions l$\leftrightarrow$r of a single residue along
dihedral angle $\phi_i$ first requires a transition along dihedral
angle $\psi$, which occurs about 10 times faster than
l$\leftrightarrow$r transitions occurring on a 1\,ns timescale. On the
other hand, left- to right-handed chiral transitions of the entire
helix, L$\leftrightarrow$R, requires individual l$\leftrightarrow$r
transition of all residues, which occur on a 100\,ns timescale.

In the following, we focus on this last tier of the
hierarchical dynamics and study the L$\rightarrow$R transition,
represented by the sum of the $\phi_i$ angles of the five inner
residues,
\begin{equation} \label{eq:Phi}
\Phi = \sum_{i=3,7} \phi_i .
\end{equation}
It was shown in Ref.\ \onlinecite{Buchenberg15} that $\Phi$ is
equivalent to the first component of a principal component
analysis of all backbone dihedral angles. \cite{Riccardi09}
Figure \ref{fig:Aib}b shows the resulting free energy landscape
$\Delta G(\Phi)$ obtained from the equilibrium MD
simulations.\cite{note3} Adopting a product-state notation of the five inner
residues, we note that the two main minima {L} = (lllll) and {R} =
(rrrrr), are connected by four intermediate states with an increasing
number of r-residues, e.g., (rllll), (rrlll,) (rrrll) and (rrrrl). As
discussed elsewhere,\cite{Biswas18} the free energy difference
($\sim 1 k_{\rm B}T$) between the states L and R is not due to
incomplete sampling but is caused by an inaccuracy of the force field
parameterization of Aib$_9$, which is not of interest here.

As discussed in Sec.\ \ref{sec:neq}, the observation of log-periodic
oscillations along $\Phi$ requires a nonequilibrium preparation and
monitoring of the system (rather than equilibrium simulations that
average over these oscillations). To this end, we extracted all
L$\rightarrow$R transitions occurring during the $8 \times 2\ \mu$s MD
simulations (see the Supplementary Material for details). Starting
when the system enters state L, monitoring the transition to state R,
and ending when it reaches R, we thus obtain 63 independent
nonequilibrium trajectories, which are used in the subsequent
analysis.
Calculating the probability distribution $P(\Phi)$ from these
trajectories, the resulting energy landscape $\Delta {\cal G}(\Phi) =
- k_{\rm B}T P(\Phi)$
is shown in Fig.\ \ref{fig:Aib}b. Since each state along
$\Delta {\cal G}(\Phi)$ serves as prerequisite for the overall
L$\rightarrow$R transition, the energy landscape shows again the
typical staircase shape, quite similar as anticipated by the 1D
model. For Aib$_9$, the first and last barriers have a height of
$\sim 1.7$ and $2.4\, k_{\rm B}T$, the three intermediate barriers of
$\sim 0.8\, k_{\rm B}T$.
In contrast to the 1D model, however, the energy landscape
$\Delta {\cal G}(\Phi)$ of Aib$_9$ represents a projection of a
high-dimensional system on a 1D coordinate. As a consequence, for
example, the states along $\Phi$ may consist of various substates
[e.g., the first intermediate state contains the conformations
(rllll), (lrlll), (llrll), (lllrl), (llllr)], which renders the
microscopic interpretation of the barriers along $\Phi$
difficult.\cite{Altis08} 

%
%
\subsection{Results}
\vspace{-4mm}

Considering the time evolution of three individual trajectories
$\Phi (t)$, Fig.\ \ref{fig:Aib}c shows the gradual climbing of the
energy landscape $\Delta {\cal G}(\Phi)$, starting from the initial
state L until the final state R is reached.  By averaging over all
trajectories, we obtain the mean position $\langle \Phi(t) \rangle$,
which is seen to raise within 100\,ns (Fig.\ \ref{fig:Aib}e). This is
in line with the timescale analysis in Fig.\ \ref{fig:Aib}d (using the
regularization parameter $\lambda_{\rm reg} = 0.5$), which reveals
peaks at $ \tau_1 \!=\! 51\,$ns, $ \tau_2 \!=\! 3.2\,$ns, and a weak
feature at $ \tau_3 \!=\! 0.4$\,ns. The resulting multiexponential fit
of $\langle \Phi(t)\rangle$ via Eq.\ (\ref{eq:Multiexp}) is found to
be in excellent agreement with the MD data (Fig.\ \ref{fig:Aib}e).

Next we construct a MSM of the system, by considering the six minima
of $\Delta {\cal G}(\Phi)$ as metastable conformational states, see
the Supplementary Material for details. Choosing a lag time of
0.5\,ns, the resulting implied timescales $t_n$ of the MSM are 74,
2.7, 1.0, 0.5, and 0.4 ns (Fig.\ \SMAib{a}), which agree qualitatively
with the peaks of the timescale analysis (Fig.\ \ref{fig:Aib}d).
Calculating the population probabilities $p_i(t)$ of the five states,
the results of MD and MSM are in excellent agreement (Fig.\
\ref{fig:Aib}f), at least within the relatively large fluctuations of
the MD data caused by finite sampling.  Moreover, the MSM calculation
of the mean position $\langle \Phi(t)\rangle$ via Eq.\
(\ref{eq:CKtest3}) matches perfectly the MD results (Fig.\
\ref{fig:Aib}e). From the time evolution of the state populations in
Fig.\ \ref{fig:Aib}f, we see a multiscale decay of the initial state L
such that intermediate states are transiently populated (within $t_4$,
$t_3$, and $t_2$), until the system relaxes in final state (within
$t_1$). The average transition times from state L to the various
intermediate states and the final state R are calculated as 0.8, 6.9,
25, 48, and 73 ns, respectively.

We finally turn to the discussion of the discrete scale invariance of
the hierarchical dynamics of Aib$_9$. Since the MSM involves
additional assumptions (such as the choice of the conformational
metastable states), we base the discussion on the timescales
determined by the maximum-entropy timescale analysis:
$ \tau_1\!=\!51\,$ns, $ \tau_2\!=\!3.2\,$ns, and
$ \tau_3\!=\!0.4\,$ns. By calculating the differences
$\log \tau_{1} - \log \tau_2 = 1.2$ and
$\log \tau_{2} - \log \tau_3 = 0.91$, we find that the logarithmic
timescales are approximately equally spaced. From this, Eq.\
(\ref{eq:logOsc3}) predicts a log-periodic power law with an a period
of $ \tau_{\rm log} \approx 1$.
On the other hand, when we fit the MD data to a log-periodic power law
[Eq.\ (\ref{eq:logOsc4})], we obtain $\tau_{\rm log} \!=\! 0.93$ for
the period (as well as $\alpha \!=\! 0.46$ for the exponent, and the
coefficients $s_a \!=\!-23 = $, $s_b \!=\! 45$, $s_c \!=\! -4.1$, and
$\varphi \!=\! -2.5$). This good agreement of theory and fit for
$\tau_{\rm log}$ is reflected in the matching of the peaks of
the log-oscillation and the main timescales of the dynamics.

When we use Eq.\ (\ref{eq:logOsc2b}) to calculate from the period
$\tau_{\rm log} = 1$ the effective roughness of the underlying energy
landscape, we obtain $\beta \Delta E =2.3$. Compared to the energy
landscape $\Delta {\cal G}(\Phi)$ in Fig.\ \ref{fig:Aib}b, this value
is smaller than the energy difference of the overall and initial
barriers, $\beta \Delta E =4$, and larger than the average barrier
height, $\beta E_b =1.3$.
This  again  demonstrates that  barrier  heights  obtained from  a  1D
projection  of the  energy  landscape in  general  cannot be  directly
associated with the true reaction rates of a multidimensional system.

%
%
\section{Discussion and conclusion}

The emergence of discrete scale invariance and the associated
phenomenon of log-periodic oscillations in protein relaxation dynamics
rests on the existence of logarithmically spaced discrete timescales of
the process. Adopting a 1D model as well as MD data of a peptide
conformational transition, here we have outlined a scenario that
explains these findings in terms of a simple hierarchical mechanism.

Consider a global structural rearrangement of a protein, which
involves the change of several inter-dependent local interactions. A
simple example is the unzipping of a $\beta$-sheet which involves the
braking of adjacent hydrogen (H) bonds. Initially, the first H-bond is
less stabilized by the $\beta$-sheet and therefore opens and closes
frequently. Once it is open, it is easier for the next H-bond to open,
which in turn facilitates the opening of the following H-bond. This
continues until the $\beta$-sheet is completely unzipped and the
resulting state is stabilized (e.g., by entropic effects or by forming
other bonds). Less likely but also possible is that some of the inner
H-bonds opens first and start the unzipping process this way. At any
rate, we find that the inter-dependence of the consecutive H-bonds
gives rise to a hierarchical mechanism, where several fast local
conformational transitions are a prerequisite for a slow global
transition to occur.  Essentially the same picture is obtained, when
we consider the global left- to right-handed transition of the peptide
helix Aib$_9$, which requires local chiral transitions of each
individual amino acid.

To describe the global transition by a 1D reaction coordinate, we
define the sum of the distances of the individual H-bonds as a
collective coordinate. Assuming that the activation energy to break a
single H-bond is given by $\beta E_B$, the scenario gives a
staircase-like free energy landscape shown in Fig.\ \ref{fig:model}b,
where the consecutive energy barriers are of similar height.  While in
an non-interacting scheme the energy barrier of all steps would be the
same, the barrier heights of the hierarchically coupled subprocesses
may vary due to the mutual interactions.  (E.g., the first and last
barrier is larger in Fig.\ \ref{fig:Aib}b.)  We note that the
characteristic staircase shape of the energy landscape and the
stabilized end states are a consequence of the hierarchical
interactions of the problem.

Considering the time evolution of the hierarchical model, we expect a
gradual climb of the consecutive energy barriers until the final state
is reached. Although the staircase-like energy landscape appears to
suggest a sequential mechanism, the process may as well occur
cooperatively. In fact, we find for both systems that successful
global transitions typically climb the energy landscape without
stopping in an intermediate state.  Employing maximum entropy-based
timescale analysis and Markov state modeling, we have shown that the
hierarchical mechanism gives rise to two or three discrete timescales
that are roughly equidistant in log-time. According to discrete scale
invariance theory, the resulting response functions exhibit a
power-law behavior that is superimposed by log-oscillations with a
period $\tau_{\rm log}$. Remarkably, these oscillations are a direct
consequence of the hierarchical model. In particular, we have shown
that the period $\tau_{\rm log}$ of the log-oscillations directly
reflects the effective roughness of the underlying energy landscape,
which in simple cases can be interpreted in terms of its barrier
heights. That is, by measuring the logarithmic period in an
ensemble-averaged experiment or MD simulation, we may conclude on the
structure of the hierarchical free energy landscape.

In ongoing work, we wish to apply the approach to the investigation of
allosteric transitions in proteins.\cite{Wodak19} For example, a joint
experimental and MD study of the structural response of a PDZ2 domain
revealed four logarithmically equidistant timescales and complex
spectroscopic and simulated time traces.\cite{Bozovic20} While the
system may provide a challenge for a discrete scale invariance
analysis, it could shed light on the elusive nature of allosteric
communication.

%
%

\subsection*{Supplementary material}
\vspace*{-4mm}
Supplementary methods including details of the timescale analysis and
the MSM, and supplementary results including additional data for the 1D
model and of Aib$_9$.

\subsection*{Acknowledgments}
\vspace*{-4mm}

The authors thank Peter Hamm, Dima Makarov, Daniel Nagel, Steffen Wolf
and Benjamin Lickert for helpful comments and discussions. This work
has been supported by the Deutsche Forschungsgemeinschaft (DFG) within
the framework of the Research Unit FOR 5099 ''Reducing complexity of
nonequilibrium'' (project No.~431945604), the High Performance and
Cloud Computing Group at the Zentrum f\"ur Datenverarbeitung of the
University of T\"ubingen, the state of Baden-W\"urttemberg through
bwHPC and the DFG through grant no INST 37/935-1 FUGG (RV bw16I016),
and the Black Forest Grid Initiative.  \vspace*{-4mm}

\subsection*{Data availability}
\vspace*{-4mm}

All data shown are available on reasonable request.

%
%

\bibliography{\bibdir/md,\bibdir/stock,new}

\begin{thebibliography}{10}
\bibitem{note3}
{Note that we deliberately refer in Fig.\ \ref{fig:Aib} to the left-hand-side
  states by \l' and 'L' and to the right-hand-side state by 'r' and 'R',
  although the common definition in the Ramachandran plot is the other way
  round}.
\end{thebibliography}

\end{document}


\title{Supplementary Material for:\\
		Log-periodic oscillations as real-time signatures
		of hierarchical dynamics in proteins}
	
	\author{Emanuel Dorbath}
	\affiliation{Biomolecular Dynamics, Institute of Physics, Albert Ludwigs University, 79104 Freiburg, Germany}
	\email{emanuel.dorbath@physik.uni-freiburg.de; stock@physik.uni-freiburg.de}
	\author{Adnan Gulzar}
	\affiliation{Biomolecular Dynamics, Institute of Physics, Albert Ludwigs University, 79104 Freiburg, Germany}
	\author{Gerhard Stock}
	\affiliation{Biomolecular Dynamics, Institute of Physics, Albert
		Ludwigs University, 79104 Freiburg, Germany}
	\email{stock@physik.uni-freiburg.de}
	\date{\today}
	
	\maketitle
	
	\baselineskip5.4mm
	
	%
	%
	\section{Theory and methods}
        \subsection{Timescale analysis}\label{SIsec:MethodsTimeScaleAnalysis}
        \vspace{-4mm}

	The time evolution of a observable $S(t)$ can be expressed by a sum of $K$ exponential response functions
	\begin{align}
		\label{eq.T.3}
		S(t) &= s_0 - \sum_{k=1,K}s_k e^{-t/\widehat \tau_k}
	\end{align}
	with the time scales $\widehat \tau_k$ and amplitudes
        $s_k$. The term $s_0$ with $\widehat \tau_0\rightarrow\infty$ yields an offset. 
	
	In the presented analysis, the amplitudes in Eq. (\ref{eq.T.3}), the so-called spectrum, are obtained from a minimization fit. Here, the minimization tool \texttt{minimize} from \texttt{scipy.optimize}\cite{scipyV2} is used with the default algorithm and a tolerance of $10^{-3}$. This is a local minimization algorithm with a high dependency on the respective starting values, which therefore must be chosen carefully. Multiple combinations of starting parameters might be reviewed to find the most appropriate one. The expression to minimize is\cite{Lorenz-Fonfria06}
	\begin{align}
		\label{eq.T.4}
		\chi^2 - \lambda_\text{reg} S_\text{ent}
	\end{align}
	with the entropy $S_\text{ent}$ and the regularization parameter $\lambda_\text{reg}$. The $\chi^2$ function is given as
	\begin{align}
		\label{eq.T.5}
		\chi^2 &= \sum_{t}^{T}\left(\hat{S}(t)-\sum_{k=0,K}s_k e^{-t/\widehat \tau_k}\right)^2,
	\end{align}
	where the data trajectory is denoted as $\hat{S}(t)$, the number of trajectory frames as $T$ and the number of fitting parameters as $M=K+1$. Note, that for the analysis the trajectory frames are converted from the usual linear spacing into logarithmic spaced frames, which heavily reduces the number of frames used. The entropy in Eq. (\ref{eq.T.4}) is defined as
	\begin{align}
		\label{eq.T.6}
		S_\text{ent} =& \sum_{k=0,K}\Bigl\{\sqrt{s_k^2+4\tilde{s}^2} -2\tilde{s} \nonumber\\
		&-s_k\log\left[\left(\sqrt{s^2_k +4\tilde{s}^2}+s_k\right)/2\tilde{s}\right] \Bigr\}
	\end{align}
	where the coefficient $\tilde{s}$ is chosen as uniform element $\tilde{s}=(\hat{S}_\text{max}-\hat{S}_\text{min})/(100 M)$, with the 'nonlinear enhancement factor' of 1/100.\cite{Lorenz-Fonfria06}\\
	
	The regularization parameter in Eq. (\ref{eq.T.4}) is an important quantity as it controls over- and underfitting and hence influences the obtained spectrum. As first step, a good regularization parameter $\lambda_\text{reg}$ can be derived using two promising schemes:
	\begin{itemize}
		\item $\chi^2$-distribution:\cite{Lorenz-Fonfria06} For a random variable which is $\chi^2$-distributed it is expected, that the $\chi^2$-value is around the number of degrees of freedom $N_\text{DOF}=T-M$. To ensure this, $\lambda_\text{reg}$ is increased until this is the case. However, with this scheme a too conservative estimation of $\lambda_\text{reg}$ might be the result, i.e., $\lambda_\text{reg}$ is to large and only the slowest of all time scales is resolved.
		\item Bayes criterion:\cite{Lorenz-Fonfria07} For this strategy, the Bayesian posterior probability $P_0(\lambda_\text{reg})\propto\lambda_\text{reg}^{N_\text{DOF}}e^{\lambda_\text{reg} S_\text{ent}-\chi^2}$ is used. The regularization parameter is chosen where the probability distribution has its maximum value. As the exponential expression is likely to result in divergences or is below numerical resolution, it is necessary to use the logarithm of $P_0\rightarrow\ln P_0=N_\text{DOF}\ln\lambda_\text{reg}-(\chi^2-\lambda_\text{reg}S_\text{ent})$ and rescale the expression $\ln P_0\rightarrow \ln P_0 /(\ln P_0^\text{max}/10)$. The rescaling enables a more strongly pronounced maximum. Finally, $\ln P_0$ is transformed back $\ln P_0\rightarrow e^{\ln P_0}$ and normalized to its maximum.
                \end{itemize}
              
\begin{figure}[htb!]
		\centering
		\includegraphics[width=0.45\linewidth,height=0.3\linewidth]{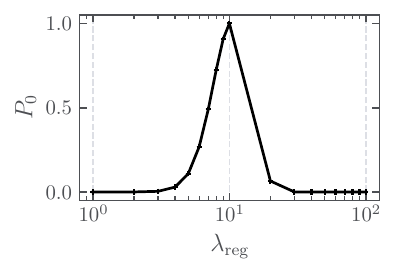}
		\includegraphics[width=0.45\linewidth,height=0.3\linewidth]{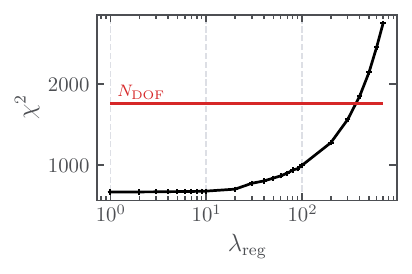}
		\caption{\setlength{\baselineskip}{4mm}
			Derivation of the optimal regularization
                        parameter $\lambda_\text{reg}$ used for the
                        time scale analysis of the 1D model. In a) the
                        Bayesian method and in b) the $\chi^2$-method is
                        shown. The red line indicates the number of
                        degrees of freedom $N_\text{DOF}$. }
		\label{SIfig:ReguParameter}
	\end{figure}

	The two methods are compared in Figure
        \ref{SIfig:ReguParameter}. Two drastically different values
        for $\lambda_\text{reg}$ are obtained, where the $\chi^2$
        method is over a order of magnitude larger than the one of the
        Bayesian probability. It is seen that latter is a better
        choice, which gives access to faster time scales while the
        $\chi^2$ one only resolves the slowest time scale.
        
	We note that the number of time scales and their range should be
        chosen with care. Reducing the fit range can result in artifacts
        at the boundaries, which is more likely to happen for the fast
        times as there is also no dominating time scale in the
        vicinity.
        Furthermore, if $S(t)$ is monotonic increasing, we can
restrict ourselves to $s_k \ge 0$, which may improve the fit.

\newpage
%
%
        \subsection{Log-periodic power law} 
        \vspace{-4mm}

\begin{figure}[htb!]
\includegraphics[width=0.75\linewidth]{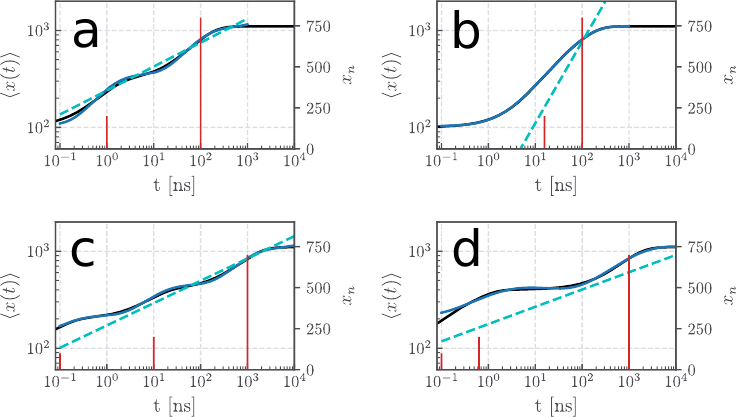}
\caption{\setlength{\baselineskip}{4mm} Fit of a multiexponential
  model $S(t) = \sum_{n} s_{n} \,e^{-t/ \tau_{n}} $ with only two
  (top) or three (bottom) main timescales $\tau_n$ and associated
  amplitudes $s_n$ to the log-periodic power law
  $ S(t) = s_a + s_b t^{\alpha} + s_c t^{\alpha} \cos ( \tfrac{2\pi}
  {\tau_{\rm log}} \log t + \varphi)$. (a) Using two timescales that
  are more than one decade apart, i.e.,
  $\log \tau_1 - \log \tau_2 \equiv \Delta_{12} =2$, we obtain
  fits showing two well-defined log-oscillations with
  $\tau_{\rm log} \approx \Delta_{12}$. (b) If the timescales are too close to each other
  ($\Delta_{12} \lesssim 1$), we effectively see only a single
  exponential term without log-oscillations.  (c) For three roughly
  logarithmically equidistant and well separated timescales, we obtain
  three well-defined log-oscillations. (d) If two of the timescales
  are too close to each other, we find only two exponential terms and
  two log-oscillations.}
\end{figure}
        
\newpage
%
%
		
	\section{Results for 1d model}
	\label{SIsec:1DModel}
	\subsection{Simulations}
	\label{SIsec:1DModelMethods}
	The 3000 trajectories are generated via simulations of the Langevin equation at $T=300$\,K and a time step of $\delta t=0.4$\,ps.\cite{Lickert21} Each trajectory starts at $x=0$ and are only stopped when $x$ becomes $\geq12.3$, i.e., the final energy barrier to \textbf{4} is crossed. This means there are not back transitions once state \textbf{4} is reached and only \textbf{1}$\rightarrow$\textbf{4} are present. However, with this scheme each trajectory has its own individual length which makes the derivation of a averaged time trace problematic. We circumvent this problem by repeating the final frame of all trajectories to match with the longest one of $\approx1.2$\,$\mu$s. As we are interested in the time trace on a logarithmic time axis, we space each trajectory logarithmically such that in each decade approximately the same number of frames are present.
	
	Fast fluctuations are reduced by Gaussian filtering, with a standard deviation of 2 frames for single trajectories and 6 frames for the averaged time trace. The latter is simply derived as arithmetic mean over the single trajectories $x_i$ while the standard deviation of the mean is calculated as unbiased estimator
	\begin{align}
		\label{eq.1DM.1}
		\sigma_{\bar{x}} = \frac{\sum_{i}^{N} |x_i-\bar{x}|^2}{N\left(N-1\right)}
	\end{align}
	where both $x_i$ and $\bar{x}$ are time dependent and $N$ are the number of produced trajectories.
	
\begin{figure}[htb!]
		\includegraphics[width=1.0\linewidth]{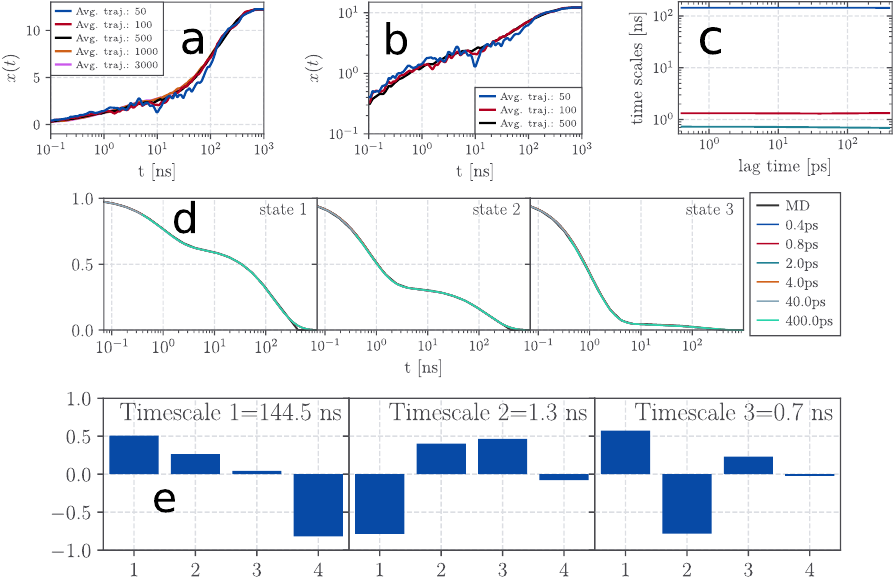}
		\caption{\setlength{\baselineskip}{4mm} (a) Mean
                  position $\langle x(t)\rangle$ of the 1d model, using
                  different numbers of trajectories. (b) double
                  logarithmic representation of $\langle x(t)\rangle$. c) Implied time scales for
                  various lag times. The time scales are almost
                  constant over several orders of magnitude verifying
                  a highly Markovian system. d) Chapman-Kolmogorov
                  test for the 3 time scales and various lag
                  times. Perfect Markovianity is verified for all lag
                  times. e) Values of the eigenvectors of the
                  transition matrix for the three implied time
                  scales. The first time scale describes the forward
                  propagation into state \textbf{4}, the second the
                  backwards transition into state \textbf{1} and the
                  third one a flux into state \textbf{2}.}
		\label{SIfig:1DModel}
	\end{figure}

	In Figure \ref{SIfig:1DModel}a, the average of the mean position $\langle x(t)\rangle$ for different numbers of trajectories is presented. The selected trajectories are smoothed after averaging as it is done for the actual analysis ($\sigma=6$ frames on a logarithmic scale). For few trajectories $N<10^2$ there are still significant and fast fluctuations visible, which might be misinterpreted as oscillations as seen in the respective log-log Figure \ref{SIfig:1DModel}b. These fluctuations vanish completely for $N>10^3$. 
	
	\subsection{Markov state model}
	\label{SIsec:1DModelMSM}
	The Markov state model is generated using a lag time of $\tau_\text{lag}=\delta t=0.4$\,ps. Each trajectory is transformed into state trajectories with the 4 states: \textbf{1} [-0.6,0.7], \textbf{2} [3.6,5.8], \textbf{3} [9.2,10.8] and \textbf{4} $>12.3$. Three implied time scales are obtained at $t_1=145$\,ns, $t_2=1.3$\,ns and $t_3=0.7$\,ns which are constant over the whole time as seen in Figure \ref{SIfig:1DModel}c. By deriving the eigenvectors of the transition matrix, the flux between the respective states for each time scale can be seen, displayed in Figure \ref{SIfig:1DModel}e. The slowest one corresponds to the \textbf{1}$\rightarrow$\textbf{4} while the second one is mostly the transition back into the initial state. The final and fastest time scale are transitions from state \textbf{1} and \textbf{3} into state \textbf{2}.
	
	The validity of the Markov state model is verified by a Chapman-Kolmogorov test. For each of the 3 states, multiple lag times $k\tau_\text{lag}$ are used to derive the both sides of the equation
	\begin{align}
		\label{eq.CK.1}
		T(k\tau_\text{lag}) = T^k(\tau_\text{lag}).
	\end{align}
	The left hand side, referred to as MD, is the transition probability estimated for a lag time $k\tau_i$, while for the right hand side the transition probability for lag time $\tau_\text{lag}$ is propagated $k$ times. Finally, a projection onto the diagonal elements is performed. As seen in Figure \ref{SIfig:1DModel}d, for both short and long lag times, a very good match with the MD is observed, verifying a high Markovianity.
	
	\subsection{Equilibrium autocorrelation function}
	\label{SIsec:1DModelAutoCorrelation}
	The autocorrelation function is derived for a single 16\,$\mu$s long equilibrium trajectory.\cite{Lickert21} An expected decay towards zero is seen which is reached latest at 100ns. A time scale analysis is performed for this correlation function giving rise to two time scales at 25\,ns and 1.3\,ns.
	\begin{figure}[htb!]
		\centering
		\includegraphics[width=0.4\linewidth]{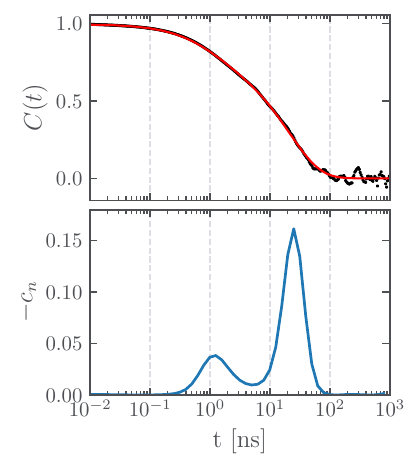}
		\caption{\setlength{\baselineskip}{4mm}
			Time scale analysis of the ACF for a single 16\,$\mu$s long trajectory with $\lambda=0.1$ and 50+1 fit parameters. In black the ACF is shown while in red the respective fit.}
		\label{SIfig:1DModelAutocorrelation}
	\end{figure}
	
%
%
	\section{Results for Aib$_9$}
	\label{SIsec:Aib9}
	\subsection{Trajectory slicing and data preparation}
	\label{SIsec:Aib9TrajectorySlicing}
	Initially, a 8x2\,$\mu$s long continuous equilibrium trajectory is generated with an effective time step of $\delta t=1$\,ps. As stated in the main text, the quantity to describe Aib$_9$ best is the cumulative angle $\Phi=\sum_{i=3}^{7}\phi_i$ over the 5 inner dihedral angles. We are interested in L$\rightarrow$R transitions and thus, it is need to slice the EQ trajectory.\\
	To this end, a Gaussian smoothing with $\sigma=2$\,ps is applied and the trajectory is transformed into a state trajectory with the two states L=[-500$^\circ$,-200$^\circ$] and R=[200$^\circ$,500$^\circ$]. All frames which do not match one of the states is set as undefined as they are irrelevant for the slicing procedure. Next, dynamical coring\cite{Nagel19} is applied with a coring time $t_\text{cor}=100$\,ps, i.e., to be counted as L/R state the state trajectory must remain uninterrupted $t_\text{cor}$ in the respective state. Finally, the trajectory is sliced from the respective first frame in L to the first one in R, thus giving the same data structure as for the 1D model. In total this gives 63 L$\rightarrow$R transitions.\\	
	
	\subsection{Markov state model}

	For the Markov State Model, the 6 states visible in $\Phi$ are used which are defined as $\pm25^\circ$ around their theoretical core: L=-250$^\circ$, rL=-150$^\circ$, rrL=-50$^\circ$, R$\ell\ell$=50$^\circ$, R$\ell$=150$^\circ$ and R=250$^\circ$. The 63 trajectories are again transformed into state trajectories, however now unmatched frames are set to the latest populated state. To enforce the L$\rightarrow$R transition, the final frame is always set to be in R and repeated often enough to be not missed by the lag time.

\begin{figure}[htb!]
		\centering
		\includegraphics[width=0.8\linewidth]{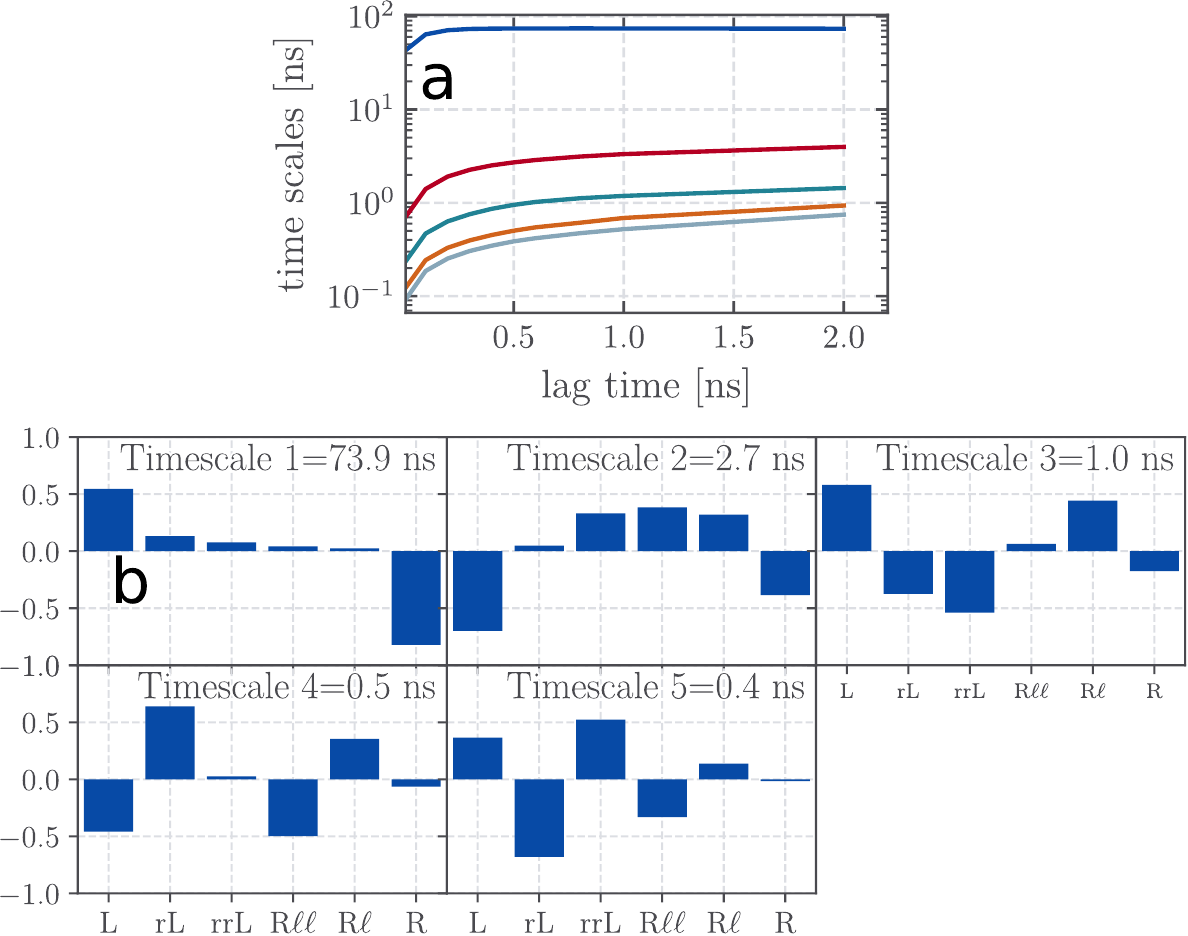}
		\caption{\setlength{\baselineskip}{4mm}
			a) Implied time scales for various lag times. The timescales become approximately constant at $\tau_\text{lag}=0.5$\,ns. b) Eigenvectors of the transition matrix.}
		\label{SIfig:Aib9MSM}
	\end{figure}

	An optimal lag time is derived at which the implied time scales become linear w.r.t. the lag rates. This is fulfilled at $\tau_\text{lag}=0.5$\,ns, see Figure \ref{SIfig:Aib9MSM}a. In total 5 implied time scales are derived with the two most important ones being at $t_1=73.9$\,ns and $t_2=2.7$\,ns. With the eigenvectors of the transition matrix, the slower one describes the full conformational change L$\rightarrow$R, while the next faster one is the result of transition of the central metastable states to the outer ones. All other time scales represent mixtures of various processes (Figure \ref{SIfig:Aib9MSM}b).\\
	
%
        %
\vspace{-8mm}        
	\bibliographystyle{\bibdir/aip+title}
\bibliography{\bibdir/stock,\bibdir/md,\bibdir/new}